\documentstyle[aps,multicol]{revtex}
\draft

\begin{document}
\title{Cotunneling through quantum dot with even number of electrons
\thanks{Contribution to LT22 (to be published in Physica B)}
}
\author{Michael Pustilnik$^{\dagger}$\cite{MP},
 Yshai Avishai$^\ddagger $, Konstantin Kikoin$^\ddagger $}
\address{$^\dagger $Danish Institute of Fundamental Metrology, \\
Anker Engelunds Vej 1, Building 307, DK-2800 Lyngby, Denmark \\
$^\ddagger $Department of Physics, Ben Gurion University, \\
Beer Sheva 84105, Israel }
\maketitle

\begin{abstract}
We study an influence of a {\it finite} magnetic field on a small spin-degenerate
quantum dot with even number of electrons, attached to metallic leads. It
is shown that, under certain conditions, the low energy physics of the
system can be described by the $S=1/2$ antiferromagnetic Kondo model.
\end{abstract}

\bigskip
\pacs{{\it Keywords:}  quantum dots; Coulomb blockade; 
cotunneling; Kondo effect}

\bigskip
\begin{multicols}{2}

Recently, the Kondo effect was observed in quantum dot systems \cite
{Experiments}, in agreement with the theoretical predictions \cite{AM}.
Quantum dots are highly tunable devices, which offer access to regimes (in
parameter space), that are virtually impossible to realize in traditional
Kondo systems. In particular, by tuning the potential at the capacitively
coupled gate electrode, one can control the number $N$ of electrons in the
dot. Away from the so-called degeneracy points, an addition or removal of an
electron to the dot costs a large charging energy $E_{c}$. Fluctuations of
the dot charge are suppressed, and $N$ is fixed. Transport in this regime
occurs by means of virtual transitions via excited states of the dot
(cotunneling). If $N$ is odd, the dot has non-zero total spin, and
cotunneling can be viewed as a magnetic exchange. An associated collective
effect manifests itself as a Kondo peak in the differential conductance at
zero bias. If $N$ is even, the total spin of the dot is zero, since all
states are occupied by pairs of electrons with opposite spins. Therefore, it
is natural to expect the parity effect: the Kondo peak is present for odd $N$%
, but not for even $N$, which has indeed been noticed in the pertinent
experiments \cite{Experiments}. Not surprisingly, most of the theoretical
attention has been focused on the quantum dots with odd number of electrons.

Yet, as we demonstrate below, dots with even $N$ may also exhibit the Kondo
effect. For this purpose we consider the following model of a
spin-degenerate quantum dot attached to a {\it single} lead \cite{transport}%
: 
\begin{equation}
{\cal H}=H_{0}+H_{d}+H_{T}.
\end{equation}
Here $H_{0}=\sum_{ks}\epsilon _{k}\psi _{ks}^{\dagger }\psi _{ks}$ describes
electrons in the lead. The Hamiltonian of the dot 
\[
H_{d}=\frac{\Delta }{2}\sum_{ps}pd_{ps}^{\dagger }d_{ps}-\frac{H}{2}%
\sum_{pss^{\prime }}\sigma _{ss^{\prime }}^{z}d_{ps}^{\dagger }d_{ps^{\prime
}}+H_{int}, 
\]
where $H_{int}=E_{c}\left( N-2\right) ^{2}$, $N=\sum_{ps}d_{ps}^{\dagger
}d_{ps}$ is the number of electrons in the dot, $p=\pm 1$ counts the energy
levels in the dot, $\Delta $ is the single particle level spacing, and $H$
is Zeeman splitting, induced by external magnetic field, and $s=\uparrow
,\downarrow $ denotes projections of spin. For simplicity, we restrict our
attention to the particle-hole symmetric case. Furthermore, we neglect all
single particle energy levels, except the two closest ones to the Fermi
level (a similar approximation is implied in the applications of the
Anderson model for dots with odd $N$ \cite{AM}). We also neglect the effect
of the magnetic field in the plane of 2DEG on conduction electrons, which is
justified as long as the bandwidth $D\sim E_{c}$ is large compared to $H$.
The tunneling Hamiltonian describes the coupling between the dot and the
lead: 
\[
H_{T}=v\sum_{ps}\psi _{s}^{\dagger }d_{ps}+{\rm {H.c}.},\;\psi _{s}=\frac{1}{%
\sqrt{L}}\sum_{k}\psi _{ks} 
\]

In weak magnetic field $H\ll \Delta $ the ground state of the dot, $%
d_{-\downarrow }^{\dagger }d_{-\uparrow }^{\dagger }\left| 0\right\rangle $,
is non-degenerate, and separated by the gap $\Delta $ from the lowest
excited states. Accordingly, the differential conductance exhibits no
structure near zero bias, but resonant peaks far away from equilibrium at $%
eV\sim \pm \Delta $ \cite{Experiments}. Qualitatively different situation
occurs if the magnetic field is tuned to the vicinity of $\Delta $: $%
H=\Delta +h,\,h\ll \Delta $. As one can easily verify, two states of the
dot, 
\begin{equation}
\left| \uparrow \right\rangle =d_{+\uparrow }^{\dagger }d_{-\uparrow
}^{\dagger }\left| 0\right\rangle \quad \,{\rm and}\,\quad \left| \downarrow
\right\rangle =d_{-\downarrow }^{\dagger }d_{-\uparrow }^{\dagger }\left|
0\right\rangle ,  \label{TLS}
\end{equation}
are almost degenerate ($E_{\downarrow }=-\Delta ,\,E_{\uparrow }=-\Delta -h$%
), and are separated by $\Delta $ from the rest of the spectrum. Clearly, if
the level spacing $\Delta $ is large, the Hilbert space of the dot can be
further truncated to the two level system (\ref{TLS}). Integration out
virtual transitions to the states with large electrostatic energy ($N=1,3$)
results in an effective low-energy Hamiltonian 
\begin{equation}
H_{eff}=H_{0}-hS^{z}+\lambda s^{z}+J\left( {\bf s}\cdot {\bf S}\right) .
\label{Heff}
\end{equation}
Here $J=2\lambda =4v^{2}/E_{c}$, $s^{\alpha }=\sum_{ss^{\prime }}\psi
_{s}^{\dagger }\left( \sigma _{ss^{\prime }}^{\alpha }/2\right) \psi
_{s^{\prime }}$ is the spin of the conduction electron at the site of the
dot, and $S$ are the usual spin 1/2 operators, acting on the states (\ref
{TLS}). Apart from the term $\lambda s^{z}$, (\ref{Heff}) is identical to
the standard form of the $S=1/2$ Kondo model. However, this term has little
influence. Indeed, it can be incorporated into the local density of states
of the conduction electrons at the site of the dot $\rho _{s}\left( \omega
\right) $. Approximating the slowly varying functions $\rho _{s}\left(
\omega \right) $ by their values at the Fermi energy, we obtain $\rho
_{s}\approx \rho \left[ 1+\left( \pi \rho _{0}\lambda /2\right) ^{2}\right]
^{-1}$, where $\rho $ is the (constant) density of states for the
Hamiltonian $H_{0}$. The remaining terms result in the Kondo contribution to
the differential conductance \cite{transport} $G(V)\propto \left( \ln
A/T_{K}\right) ^{-2}$, where $A=\max \left\{ \left| eV\right| ,\left|
h\right| ,T\right\} $, and the Kondo temperature is defined as $T_{K}\sim
E_{c}\left( \rho J\right) ^{1/2}\exp {\left( -1/\rho J\right) }$.

To conclude, we predict in this paper that the quantum dots with even number
of electrons may exhibit Kondo effect in a {\it finite} magnetic field. This
is possible since the energy gap for spin excitations in quantum dots is
equal to a single particle level spacing, which is small compared to the charging
energy. More realistic model should take into account an assymetry of the
tunneling amplitudes to the two energy levels, participating in the effect.
This will be discussed elsewhere \cite{we}.

We are grateful to Karsten Flensberg for valuable discussions. We
acknowledge support of the European Commission through the Contract
SMT4-CT98-9030 (MP), the Kreitman fellowship of the Ben Gurion University
(MP) and the Israeli National Fund center of excellence program(YA). One of
us (MP) acknowledges the hospitality of the Oersted Laboratory of Niels Bohr
Institute.

\end{multicols} 


\begin{references}
\bibitem[@]{MP}E-mail address: misha@fys.ku.dk
\bibitem{Experiments}  D. Goldhaber-Gordon et al, Phys. Rev. Lett. {\bf 81},
5225 (1998); D. Goldhaber-Gordon et al, Nature {\bf 391}, 156 (1998); S. M.
Cronenwett, T. H. Oosterkamp, and L. P. Kouwenhoven, Science {\bf 281}, 540
(1998); J. Schmid, J. Weis, K. Eberl, and K. von Klitzing, Physica B {\bf %
256-258}, 182 (1998).

\bibitem{AM}  L. I. Glazman and M. E. Raikh, JETP Lett. {\bf 47}, 452
(1988); T. K. Ng and P. A. Lee, Phys. Rev. Lett. {\bf 61}, 1768 (1988); S.
Hershfield, J. H. Davies, and J. W. Wilkins, Phys. Rev. Lett. {\bf 67}, 3720
(1991); Y. Meir, N. S. Wingreen, and P. A. Lee, Phys. Rev. Lett. {\bf 70},
2601 (1993).

\bibitem{transport}  The substitution $\psi _{s}=2^{-1/2}\sum_{\alpha
}e^{i\alpha eVt/2}\psi _{\alpha s}$ generalizes the model to the case when
there are two leads and source-drain voltage, necessary for computation of
the tunneling current. Here $\alpha =+1/-1$ for the right/left lead
accordingly.

\bibitem{we}  M. Pustilnik, K. Kikoin, and Y. Avishai, to be published.

\end{references}
\end{document}